\documentclass[%
 aps,
prb,
superscriptaddress,
 amsmath,amssymb,
 preprint,%
]{revtex4-2}

\usepackage{graphicx}% Include figure files
\usepackage{dcolumn}% Align table columns on decimal point
\usepackage{bm}% bold math
\usepackage{hyperref}
\usepackage[utf8]{inputenc}
\usepackage[T1]{fontenc}
\usepackage{mathptmx}

\begin{document}

\preprint{}

\title[Wurtzite vs rock-salt MnSe epitaxy: electronic and altermagnetic properties]{Wurtzite vs rock-salt MnSe epitaxy: electronic and altermagnetic properties}
% Force line breaks with \\

\author{Michał J. Grzybowski}
 \email{Michal.Grzybowski@fuw.edu.pl}
 %\altaffiliation[Currently at: ]{Eindhoven University of Technology, 5600 MB Eindhoven, the Netherlands}
 \affiliation{Faculty of Physics, University of Warsaw, ul. Pasteura 5, 02-093 Warsaw, Poland}
\author{Carmine Autieri}%
 \affiliation{International Research Centre Magtop, Institute of Physics, Polish Academy of Sciences, Aleja Lotników 32/46, PL-02668 Warsaw, Poland}
\author{Jarosław Domagała}
 \affiliation{Institute of Physics, Polish Academy of Sciences, Aleja Lotnikow 32/46, PL-02668, Warsaw, Poland}
\author{Cezary~Krasucki}
 \affiliation{Faculty of Physics, University of Warsaw, ul. Pasteura 5, 02-093 Warsaw, Poland}
 \affiliation{Institute of Physics, Polish Academy of Sciences, Aleja Lotnikow 32/46, PL-02668, Warsaw, Poland}
\author{Anna Kaleta}
  \affiliation{Institute of Physics, Polish Academy of Sciences, Aleja Lotnikow 32/46, PL-02668, Warsaw, Poland}
\author{Sławomir Kret}
 \affiliation{Institute of Physics, Polish Academy of Sciences, Aleja Lotnikow 32/46, PL-02668, Warsaw, Poland}
\author{Katarzyna Gas}
 \affiliation{Institute of Physics, Polish Academy of Sciences, Aleja Lotnikow 32/46, PL-02668, Warsaw, Poland}
 \affiliation{Center for Science and Innovation in Spintronics, Tohoku University, Sendai 980-8577, Japan}
\author{Maciej Sawicki}
 \affiliation{Institute of Physics, Polish Academy of Sciences, Aleja Lotnikow 32/46, PL-02668, Warsaw, Poland}
\author{Rafał Bożek}
  \affiliation{Faculty of Physics, University of Warsaw, ul. Pasteura 5, 02-093 Warsaw, Poland}
\author{Jan~Suffczyński}
  \affiliation{Faculty of Physics, University of Warsaw, ul. Pasteura 5, 02-093 Warsaw, Poland}
\author{Wojciech Pacuski}
  \affiliation{Faculty of Physics, University of Warsaw, ul. Pasteura 5, 02-093 Warsaw, Poland}

 %\homepage{http://www.Second.institution.edu/~Charlie.Author.}
%\affiliation{%
%School of Physics and Astronomy, University of Nottingham, United Kingdom%\\This line break forced% with \\
%}%

\date{\today}% It is always \today, today,

\begin{abstract}
Newly discovered altermagnets are magnetic materials exhibiting both compensated magnetic order, similar to antiferromagnets, and simultaneous non-relativistic spin-splitting of the bands, akin to ferromagnets. This characteristic arises from the specific symmetry operations that connect the spin sublattices. In this report, we show with {\it ab initio} calculations that the semiconductive MnSe exhibits altermagnetic spin-splitting in the wurtzite phase as well as a critical temperature well above room temperature. It is the first material from such space group identified to possess altermagnetic properties.
Furthermore, we demonstrate experimentally through structural characterization techniques that it is possible to obtain thin films of both the intriguing wurtzite phase of MnSe and the more common rock-salt MnSe using molecular beam epitaxy on GaAs substrates. The choice of buffer layers plays a crucial role in determining the resulting phase and consequently extends the array of materials available for the physics of altermagnetism.
\end{abstract}

\maketitle

\section{Introduction}

Intensive research on the physics of antiferromagnets in recent years led to the identification of materials that combine the properties of antiferromagnets (AFM) such as magnetization compensation together with features of typical ferromagnets like spin-split band structures. Such coexistence of effects was shown to originate from the presence of rotation symmetry operations and the absence of translation symmetry operations connecting different spin-sublattices in a generalized symmetry formalism and allowed for a definition of altermagnets \cite{Smejkal2022, Smejkal2022b} which are distinct from both ferromagnets and antiferromagnets. Altermagnets are very appealing due to their potential applications, for example they can exhibit outstanding charge-spin conversion efficiency \cite{Gonzalez-Hernandez2021,Bose2022,Bai2022,Karube2022}, large giant or tunneling magnetoresistance values \cite{Smejkal2022_GMR} together with THz spin dynamics \cite{Smejkal2022}.
Numerous materials fulfill the symmetry criteria to be altermagnets ranging from insulators to metals \cite{Smejkal2022}. Among them, hexagonal MnTe is an important chalcogenide semiconductor recently shown to exhibit anomalous Hall effect \cite{GonzalezBetancourt2023}. This motivated us to search for altermagnetism in various crystal structures of manganese-based chalcogenides.

In this work, we investigate MnSe, which can form a variety of phases, although the most common form is the rock-salt structure. We show the possibility to stabilize the epitaxial growth of two different phases of thin film MnSe - rock-salt (RS) or wurtzite (WZ), depending on the proper choice of growth conditions with the use of the same MBE technique and GaAs substrates. During all of these processes, we observe sharp and bright reflection high-energy electron diffraction (RHEED) patterns being a sign of high-quality 2D growth. Of particular interest is the wurtzite phase that had been shown to be unstable in bulk form \cite{Baroni1938} and thus rare. It was observed either as a minor impurity phase or in colloidal nanoparticles \cite{Sines2010} but not for epitaxial thin films. We explore the structural properties and quality of thin films of MnSe by X-ray diffraction (XRD), transmission electron microscopy (TEM), energy-dispersive X-ray spectrometry (EDX), atomic force microscope as well as Raman spectroscopy and superconducting quantum interference device (SQUID) magnetometry for the thicker layers of rock-salt MnSe.

Moreover, we perform {\it ab initio} calculations to explore electronic band structure and we find altermagnetic spin-splitting in the wurtzite MnSe together with the critical temperature above room temperature. The altermagnetism depends on the magnetic space groups, therefore as a consequence also on the Wyckoff positions of the magnetic atoms, magnetic order and dimensionality \cite{Smejkal2022, Cuono23orbital, Fakhredine23}. Several compounds belonging to the hexagonal space group no. 194 (NiAs phase) show altermagnetism as the MnTe and EuIn$_2$As$_2$ compounds \cite{Smejkal2022,Cuono23EuCd2As2,GonzalezBetancourt2023}.
Going to the hexagonal space group no. 186 (wurtzite phase), MnSe is the first altermagnetic compound proposed within this space group to our knowledge.

%be possible due was shown to coexist have 
Historically, the interplay of different phases has been observed very early for the bulk MnSe. Although, the most thermodynamically stable form of MnSe is the rock-salt structure MnSe \cite{Schlesinger1998,D1NR00822F}, different crystalline phases were believed to transform between each other upon cooling and heating cycles, which was considered as a likely reason for a strong thermal hysteretic behavior of its magnetic properties \cite{Lindsay1951}. Alternative explanations have been provided \cite{Ito1978} but more recent research proves the existence of admixture of hexagonal NiAs-type MnSe in lower temperatures \cite{Huang2019, Efrem2004} and correlates it with the existence of hysteresis. Remarkably, high pressure may induce transformation to the orthogonal phase and superconductivity \cite{Hung2021, Man2022}. Epitaxial growth of MnSe on (100) GaAs is known to result in a rock-salt structure \cite{Kim1998} except for very thin layers \cite{Kim1998, Heimbrodt1993, Chang1987}. Successful efforts of stabilizing the zinc-blende structure in MBE were undertaken by the growth of superlattices with ZnSe on (100) GaAs \cite{Kolodziejski1986, Heimbrodt1993, Ishibe2000}. The growth of MnSe on a $c-$plane sapphire \cite{Kucharek2019} or NbSe$_2$ \cite{Aapro2021} substrates have also been attempted. The latter can result in a monolayer of unusual ordering. Computations confirm that the choice of the substrate and resulting strain can have a tremendous effect on the magnetic properties of this material \cite{Nakamura2007}. 

\section{Experimental results and discussion}
\subsection{Rock-salt MnSe}
All MnSe layers considered in this work were grown in II–VI MBE growth chamber equipped with standard Knudsen effusion cells for both Se and Mn sources. We start from the general notice that the growth of MnSe directly on GaAs without any buffer layers results in a rock-salt structure of the former. For GaAs (100) with $2^\circ$ offcut substrates, the RHEED image gradually transforms from the 2D streaky to the 3D spotted pattern. The change of the distance between the lines being a sign of diminishing lattice constant can be observed. Significant lattice mismatch is expected to promote the formation of defects. The crystal structure quality was studied by XRD and reflections corresponding to (002), (004) and (006) rock-salt MnSe could be observed in $2\theta/\omega$ scans. The width of the (004) reflection was studied as a function of the temperature of the growth. Although a slightly narrower peak has been observed for the temperature of $T=350^\circ$C, we choose the temperature of $T=300^\circ$C as optimal for the growth of all samples considered in this work. This is because elevated temperatures preclude the growth of numerous II-VI buffer layers.

\begin{figure}[h!]
\centering
  \includegraphics[width=0.75\linewidth]{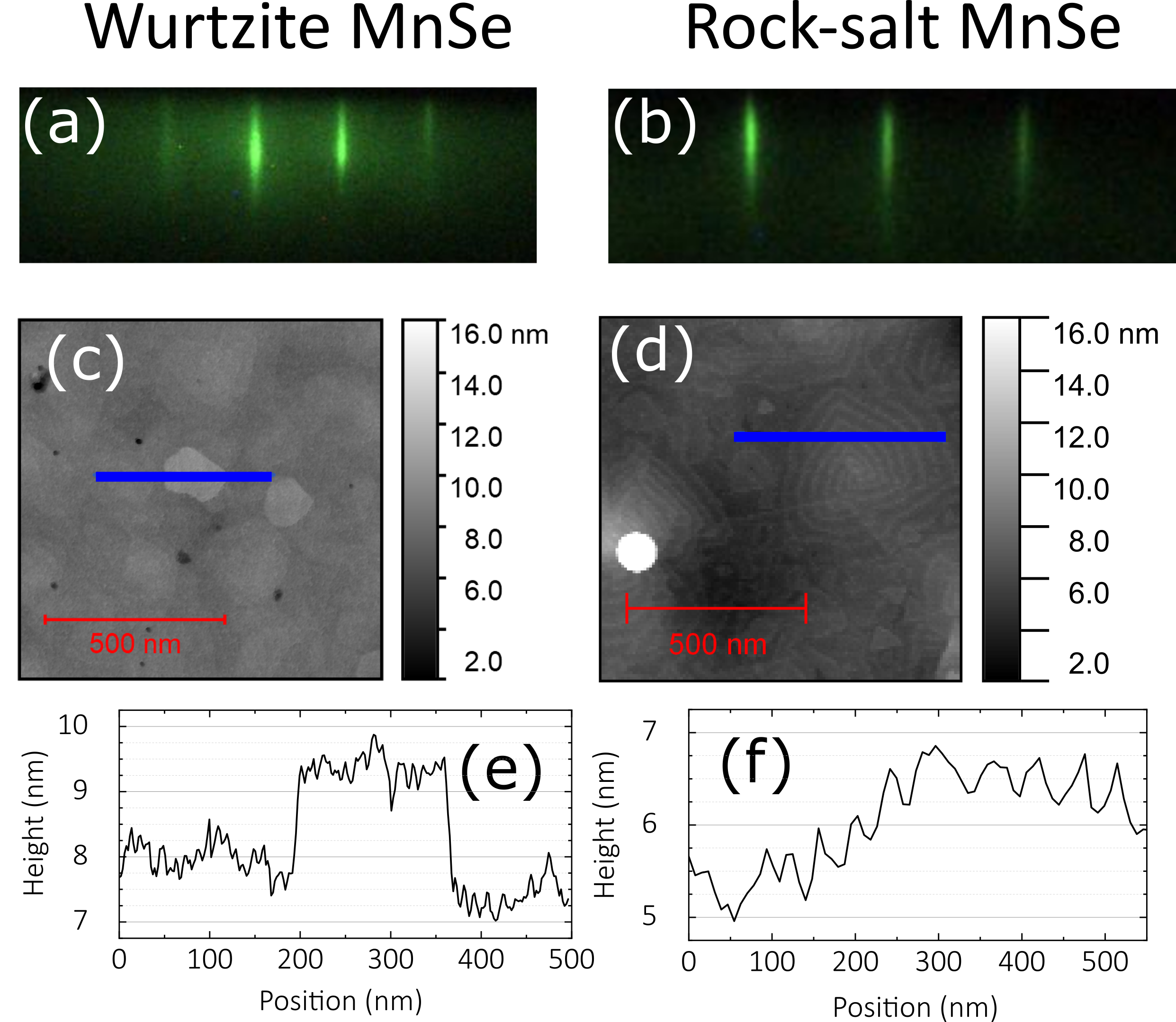}
  \caption{Comparison of structural characteristics of the top surface of wurtzite MnSe (left column, sample WZ3) and rock-salt MnSe (right column, sample RS1). RHEED images (a)-(b) taken just at the end of the growth, AFM topography scans (c)-(d) and profiles along the blue lines (e)-(f).}
  \label{fig:rheedafm111}
\end{figure}

\begin{figure}[h!]
\centering
  \includegraphics[width=0.75\linewidth]{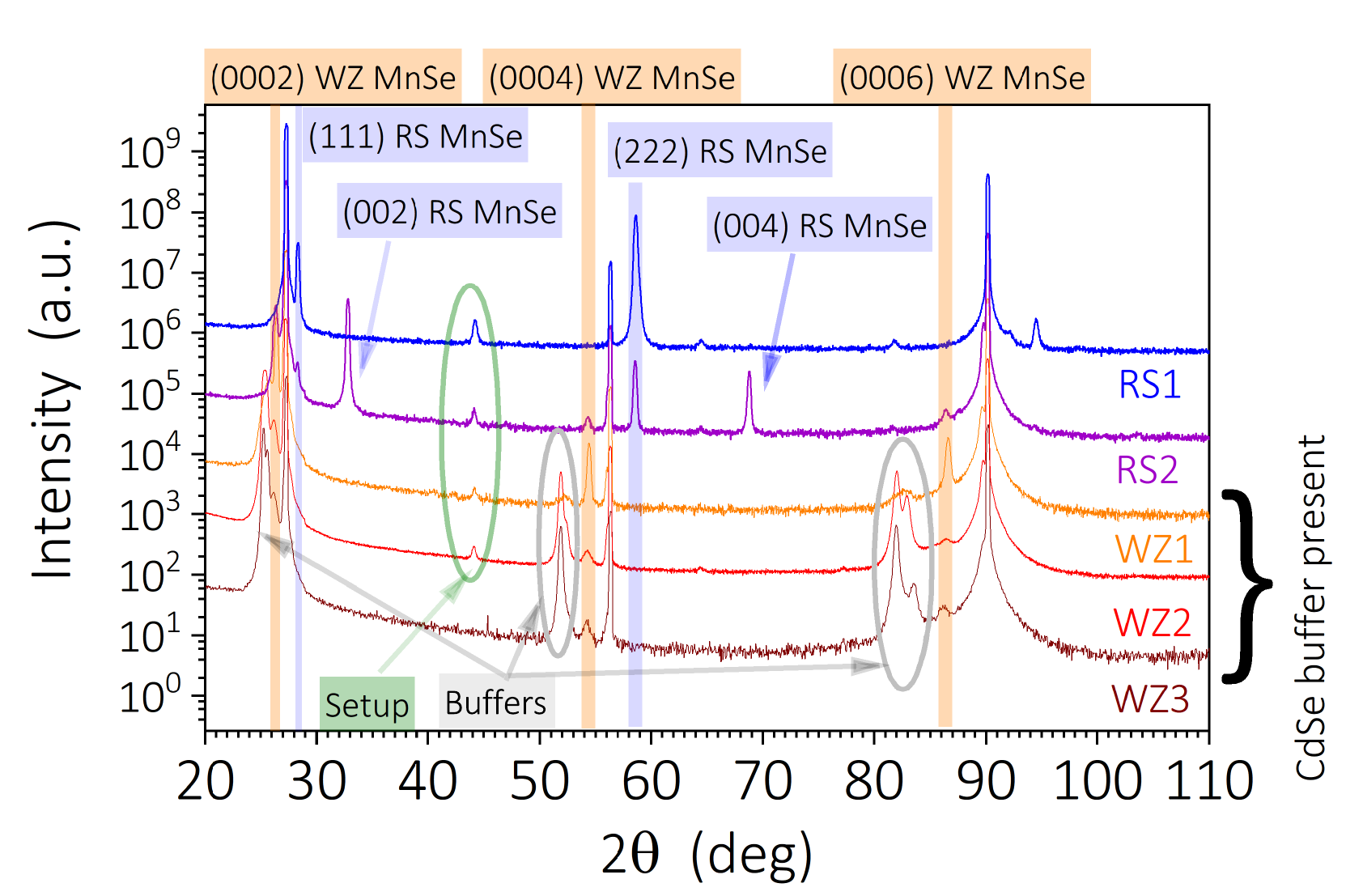}
  \caption{Comparison of $2\theta/\omega$ scans for structures with top MnSe (111) thin-film layers evidencing an emergence of wurtzite phase when CdSe buffer is present.}
  \label{fig:xrdrswz}
\end{figure}

In contrast to other works considering the growth of MnSe by MBE \cite{Kim1998, Heimbrodt1993, Chang1987, Kolodziejski1986, Ishibe2000} we also attempted the use of GaAs substrates with (111)B orientation. Remarkably, such a procedure results in the 2D RHEED pattern that remains in the form of bright and sharp lines up to the thickness of hundreds of nanometers. The surface of the $200\,$nm layer of MnSe (sample RS1) studied by atomic force microscopy exhibits subnanometer roughness measured over hundreds of nanometers paths (Fig.~\ref{fig:rheedafm111}(d)), presence of terraces of triangular shape, screw dislocation patterns and occasional large surface defects of a round shape. RMS roughness parameter calculated for a random $500\,$nm square region without large round surface defect is at the order of 0.6 nm. In XRD patterns, reflections corresponding to (111), (222) rock-salt phase of MnSe are clearly visible in $2\theta/\omega$ scans (blue line in Fig.~\ref{fig:xrdrswz}). We observe an almost fully relaxed unit cell (supplementary materials). Moreover, the sample exhibits 6-fold symmetry with respect to the <111> unlike the GaAs substrate, which displays 3-fold symmetry (supplementary materials). It is most likely the sign of crystal twinnings that arise due to the high symmetry of the plane during the growth and $0\,^\circ$ substrate offcut. It can be likely mitigated with the use of substrates with significant offcut with respect to the (111) plane.

\begin{figure}[h]
\centering
  \includegraphics[width=0.75\linewidth]{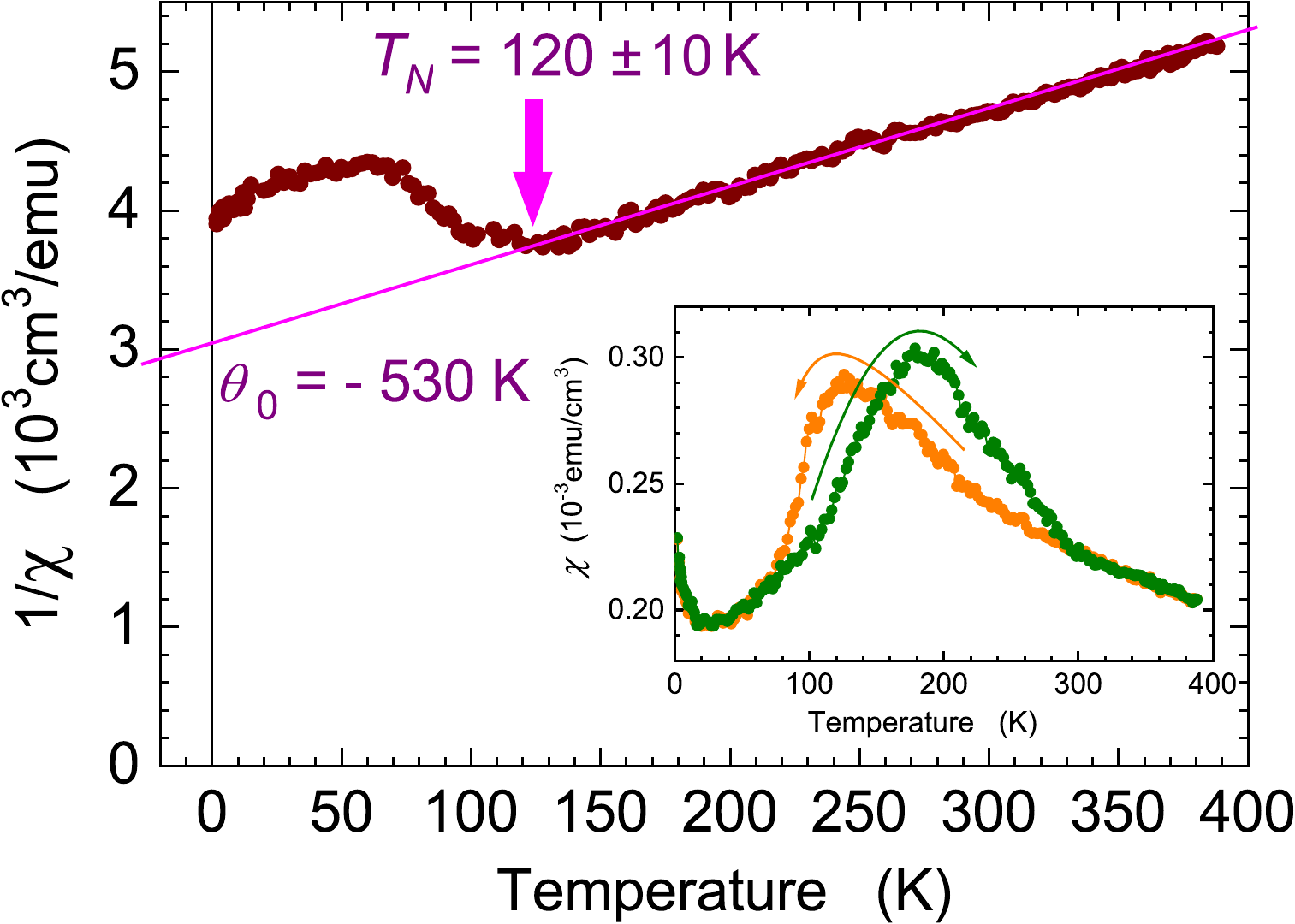}
  \caption{(Solid symbols) Inverse of magnetic susceptibility, $\chi$, of rock-salt MnSe, sample RS1, established at  $\mu_0 H  = 4$~T.  The solid line represents the Curie-Weiss behavior yielding the Curie-Weiss temperature $\Theta_{\text{CW}} = -530 \pm 30$~K and the magnitude of the spin moment $S = 2.4 \pm 0.2$. The thick arrow marks the Neel temperature $T_\text{N}$ of this compound. Inset: Thermal hysteresis of $\chi$, characteristic for rock-salt MnSe, evidenced at external field of $\mu_0 H  = 1$~T. }
  \label{fig:squid1}
\end{figure}

Magnetic measurements of the rock-salt RS1 sample were done in a SQUID-based magnetometer MPMS XL of a Quantum Design (San Diego, CA, USA). We followed strictly the code appropriate for measurements for thin magnetic layers of minute magnetic signals deposited on solid state bulky substrates \cite{Sawicki2011}. The sample was stabilized with the help of a strongly ethanol-diluted GE-varnish on a sample holder made from a length of about $1.8\,$mm wide Si stick. To correctly establish the signal specific to the thin MnSe layer a reference measurements were taken in the same temperature and magnetic field ranges for an otherwise identical piece of GaAs substrate mounted in a similar manner on the same sample holder. This extra effort is especially important since commercial solid state substrates exhibit a sizable temperature-dependent response \cite{Ney2006, Gas2022}, which can be easily comparable to that of the investigated layers, antiferromagnets in particular, and so it needs to be subtracted accordingly with adequate care. The results of our $T$-dependent studies are summarized in Fig.~\ref{fig:squid1}. In accordance with the finding of ref. \cite{Huang2019} strong thermal hystereses are seen between 70 and 300 K, as exemplified in the inset to Fig.~\ref{fig:squid1}. The magnitudes of these anomalies (in comparison to the magnetic susceptibility $\chi(T)$ at $T > 300\,$K) decrease substantially at stronger magnetic fields $H$. So, in order to establish the N\'eel temperature $T_\text{N} $ of rock-salt MnSe we limit our analysis here to results obtained at relatively high magnetic fields. The inverse of $\chi(T)$ established for $\mu_0 H  = 4$~T  is given in the main panel of Fig.~\ref{fig:squid1}. It follows a sufficiently straight line at elevated temperatures to allow us to apply the time-honored Curie-Weiss (CW) approach. The abscissa and slope of this line yield the Curie-Weiss temperature $\Theta_{\text{CW}} = -530 \pm 30$~K and the magnitude of the spin moment $S = 2.4\pm0.2$, respectively. The latter value confirms the expected $3d^5$ configuration of Mn ions in rock-salt MnSe. The experimental points follow the CW law only above about 120 K. Below, the experimental points clearly depart from the linear dependency. We assign this temperature to $T_\text{N} = 120\pm10\,$K of rock-salt MnSe. We mark this point by a thick arrow in Fig.~\ref{fig:squid1}. This is the magnitude of $T_\text{N}$ as established previously from heat capacity measurements \cite{Huang2019}. 

\begin{figure}[h]
\centering
  \includegraphics[width=0.75\linewidth]{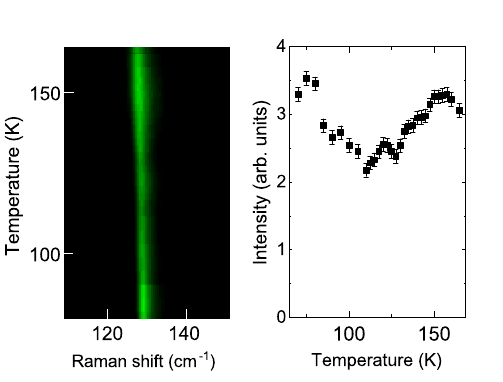}
  \caption{(a) Raman spectra of rock-salt MnSe layer (sample RS3) as a function of temperature. (b) Intensity of  Raman transition at around 130 cm$^{-1}$ as a function of temperature exhibits a clear maximum in the vicinity of the N\'eel temperature.}
  \label{fig:Raman}
\end{figure}

For Raman spectroscopy measurements the sample is placed in a helium flow cryostat. The spectra are excited with the use of 532~nm laser and the signal is collected with a microscope objective of numerical aperture NA = 0.65. Figure \ref{fig:Raman} (a) shows Raman spectra in the spectral vicinity of the transition at around 130~cm$^{-1}$ recorded for consecutive temperatures increasing from 70 K to 170 K. The transition is attributed to the excitation of the LO(X) phonon in rock-salt MnSe \cite{Popovic:PRB2006}. The intensity of the transition is presented in Fig. \ref{fig:Raman} (b). It exhibits a minimum at around 120~K, corresponding to N\'eel temperature in this material. Observed decrease of the Raman transition intensity at around T$_N$ confirms that phonon excitations provide a probe for the magnetic phase transitions in the studied system. This can be traced back to a contribution of an exchange type, apart from ordinary Coulomb-type interaction, between Mn cations possessing non-vanishing spin moment. Similar effects have been previously observed mostly in 2D layered AFMs, where phonon interactions are much stronger \cite{Vaclavkova:2DMat2020}.

Next, we describe the growth of MnSe on the buffer layers deposited on GaAs (111)B (see the full list of the investigated samples in Table~\ref{tab}). One of the commonly used buffers for II-VI semiconductors is ZnSe. However, for the growth of ZnSe in (111) orientation, the RHEED pattern evolves into a spotty 3D image precluding the growth of thick layers of high quality. For 60$\,$nm ZnSe buffer, the subsequent layer of MnSe (sample RS2) exhibits rock-salt character but with a mixture of <111> and <001> orientation and a likely admixture of wurtzite phase (purple line in Fig.~\ref{fig:xrdrswz}).

\begin{table*}[tb]
\small
\caption{\label{tab:1}The list of investigated MBE-grown samples containing rock-salt (RS) or wurtzite (WZ) MnSe as a top layer. All of the listed samples were grown on GaAs (111)B substrates. }
\label{tab}
\begin{tabular*}{\textwidth}{@{\extracolsep{\fill}}llll}
\hline
\textbf{Sample number} & \textbf{Sample strucuture (thickness in nm)} & \textbf{Phase of MnSe}\\  
\hline
\textbf{RS1} \scriptsize{(UW1932)} & \textbf{MnSe (200)} & RS \\
\hline
\textbf{RS2} \scriptsize{(UW1933)} & ZnSe (60) | \textbf{MnSe (200)}  & RS/WZ \\
\hline
\textbf{RS3} \scriptsize{(UW2097)} & \textbf{MnSe (2500)} & RS \\
\hline
\textbf{WZ1} \scriptsize{(UW1947)} & ZnSe (60) | CdSe (10) | \textbf{MnSe (60)}  & WZ \\
\hline
\textbf{WZ2} \scriptsize{(UW1961)} & ZnSe (70) | ZnTe (50) | (Cd,Mg)Se (50) | CdSe (10) | \textbf{MnSe (20)}  & WZ \\
\hline
\textbf{WZ3} \scriptsize{(UW1994)} & ZnSe (70) | ZnTe (50) | (Cd,Mg)Se (50) | CdSe (20) | \textbf{MnSe (20)} & WZ \\
\hline
\end{tabular*}
\end{table*}

%\begin{figure*}
% \centering
% \includegraphics[height=3cm]{example2}
% \caption{An image from the \textit{Physical Chemistry Chemical Physics} cover gallery, set as a two-column figure.}
% \label{fgr:example2col}
%\end{figure*}

\begin{figure*}[h]
\centering
\includegraphics[width=1\linewidth]{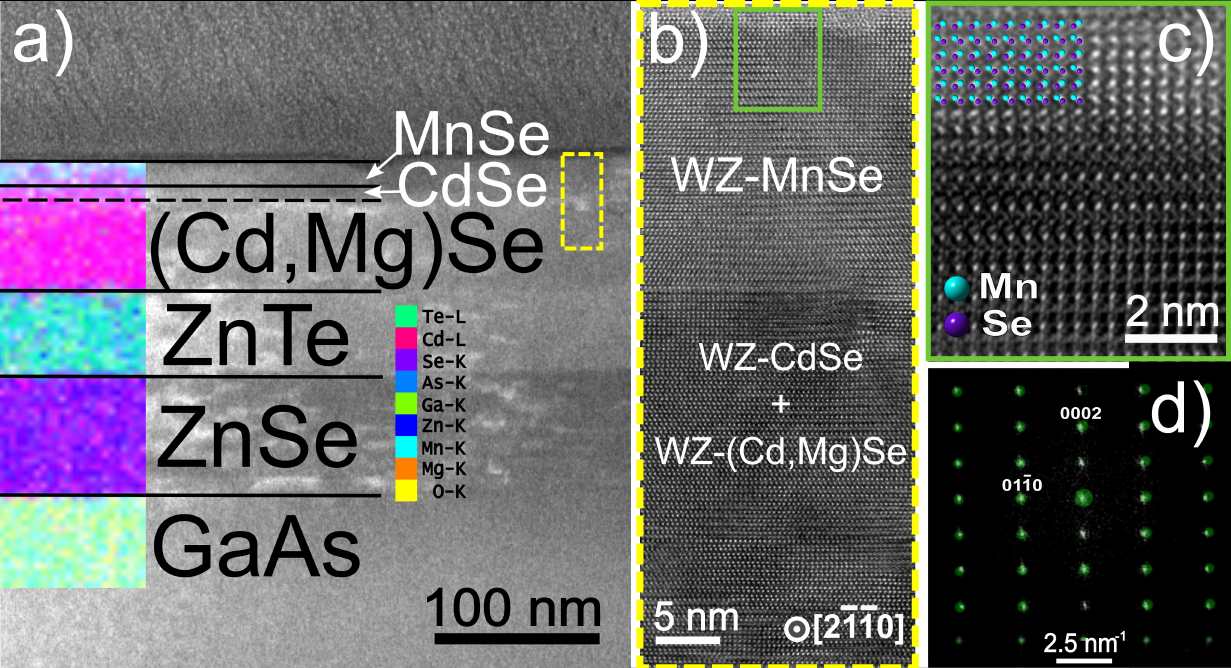}
\caption{(a) Scanning Transmission Electron Microscopy (STEM) image of sample WZ2 cross-section with energy-dispersive X-ray spectroscopy (EDX) color-maps showing GaAs substrate, ZnSe and ZnTe buffer layers, as well as (Cd,Mg)Se, CdSe, MnSe top layer; (b) High-resolution TEM (HRTEM) image of yellow rectangle area indicated in panel a, presenting wurtzite MnSe and wurtzite CdSe/(Cd,Mg)Se layers being in perfect epitaxial relationship recorded along the $[2\overline{1}\overline{1}0]$ zone axis. 
(c) Magnified area from panel b (green rectangle) with atom-ball model of wurtzite MnSe structure (visualized in $[2 \overline{1} \overline{1} 0]$ zone axis) superimposed on HRTEM image. (d) Kinematic electron diffraction simulated for wurtzite MnSe (green spots) perfectly overlapping 2D-Fast Fourier Transform (FFT) of the image in panel c (grayscale image in the background).}
\label{fig:tem}
\end{figure*}

\subsection{Wurtzite MnSe}
CdSe can be grown on GaAs (111) in a wurtzite phase \cite{Grun1994,Goppert1998,Matsumura2000}. This holds true even if additional layers of ZnSe | ZnTe are introduced in between (Fig.~\ref{fig:tem}). Remarkably, we observe that for every sample in our study in which the MnSe layer is grown on CdSe, MnSe adopts the wurtzite phase (see the list of the studied samples in Table~\ref{tab:1}). This is verified by XRD and TEM. The reflections corresponding to the (0002) and (0004) and (0006) wurtzite phase of MnSe can be seen (Fig.~\ref{fig:xrdrswz}). Due to the thin film character of the wurtzite MnSe layers (unlike the rock-salt MnSe) we focus on the detailed TEM analysis for this phase rather than bulk-sensitive techniques that are difficult to apply here such as Raman spectroscopy and SQUID magnetometry.

%Transmission electron microscopy (Fig.~\ref{fig:tem}) shows the existence of a well-ordered crystalline phase and electron diffraction confirms the wurtzite structure (\textbf{AK oraz SK attention needed}) in the WZ2 structure.
Transmission Electron Microscopy analysis in Fig.~\ref{fig:tem} shows results for the cross-section of sample WZ2 (prepared with focused ion beam), revealing its chemical composition (Fig.~\ref{fig:tem}(a) as well as detailed analysis of the crystal structure of MnSe layer (Fig.~\ref{fig:tem}(b),(c),(d)). Color-coded energy-dispersive X-ray spectroscopy (EDX) maps in Fig.~\ref{fig:tem}(a) confirm that the heterostructure outcome is in accordance with MBE specifications (see Table~1). Comprehensive structural studies on MnSe phase identification are accomplished via high-resolution TEM (HRTEM) imaging in correlation with Fast Fourier Transform (FFT) analysis (electron diffraction analog), which unequivocally confirms the wurtzite phase of MnSe. Clear evidence is presented in Fig.~\ref{fig:tem}(c) – wurtzite MnSe ball model (in $[2\overline{1}\overline{1}0]$ zone axis) is superimposed on the HRTEM image in its thinnest area (top part) as well in Fig.~\ref{fig:tem}(d), where simulated kinematic electron diffraction aligns with FFT image.

RHEED pattern of the surface of wurtzite MnSe consists of sharp and bright lines throughout the growth of the layers of at least $60\,$nm thickness (left column of Fig.~\ref{fig:rheedafm111} shows data for WZ3). The surface measured by AFM is subnanometer flat over hundreds of nm distances as indicated in Fig.~\ref{fig:rheedafm111}(c)~and (e) and RMS roughness parameter calculated for a random $500\,$nm square region without large round surface defect is at the order of 1.5 nm.

The complex set of buffer layers beneath wurtzite CdSe, introduced for some of the studied samples, was motivated by the analogous experimental protocols that had been developed and successfully tested for high-quality CdSe quantum wells grown on GaAs (001) oriented substrates\cite{Sawicki2019,Sawicki2020}. Therefore, we notice that some of our samples should form an interesting asymmetric quantum well that can be studied optically. The limitations of quality of ZnSe growth in (111) orientation and simultaneous good quality RHEED pattern for CdSe in the same samples together with the universal observation of wurtzite MnSe on CdSe within the studied set of samples suggest that the crucial factor to obtain wurtzite MnSe is to proceed the deposition on wurtzite CdSe.

\section{First Principles Calculations}

\begin{figure}[h]
\centering
\includegraphics[angle=270,scale=0.34]{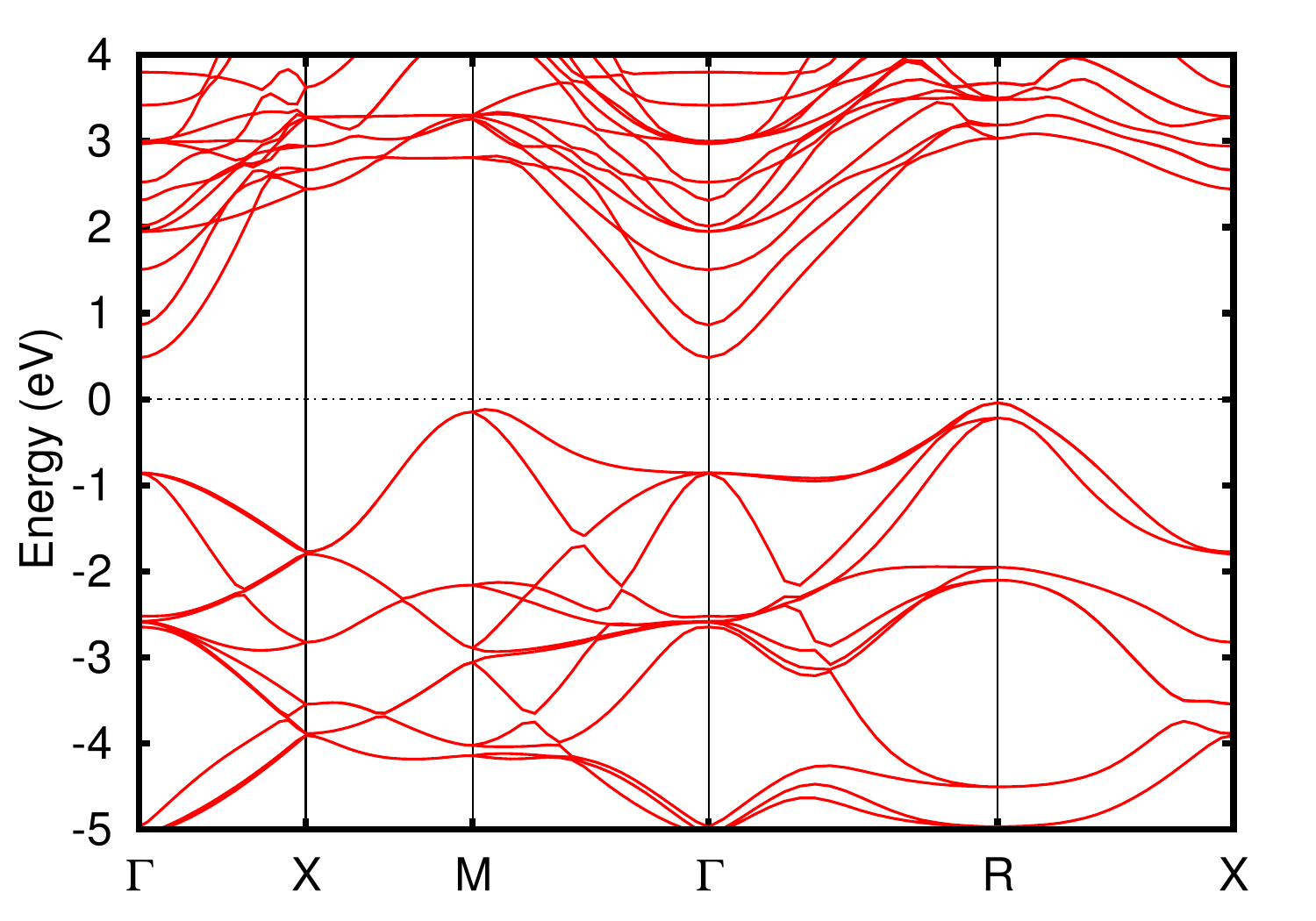}
\includegraphics[angle=270,scale=0.34]{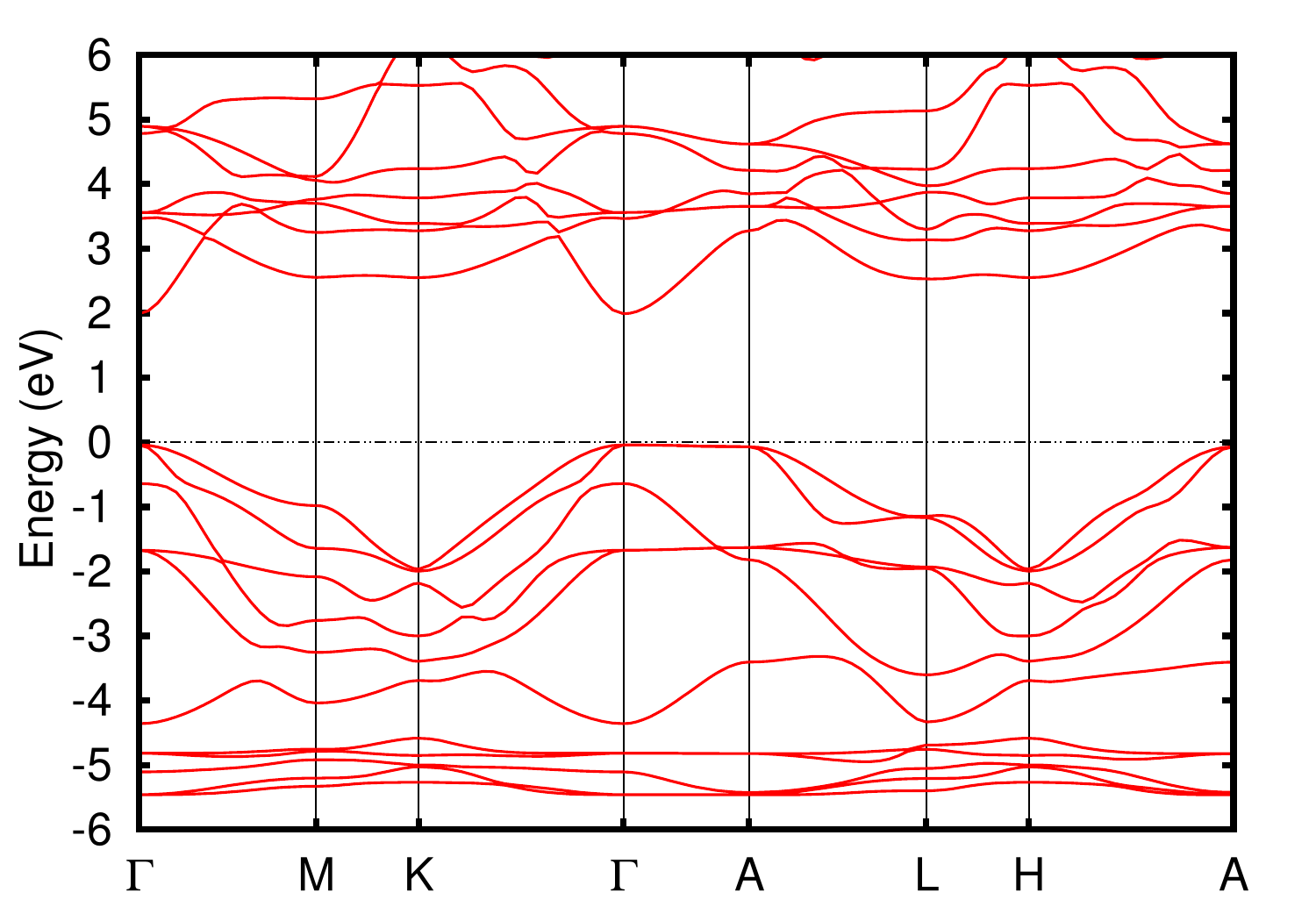}
\caption{(a) Magnetic band structure of MnSe in the rock salt phase (space group no. 225). (b) Magnetic band structure of MnSe in the wurtzite phase (space group no. 186). The Fermi level is set to zero. In both cases, we have used the value of U=6 eV.}
\label{FigureTheory1}
\end{figure}

\begin{figure}[h]
\centering
\includegraphics[angle=270,scale=0.34]{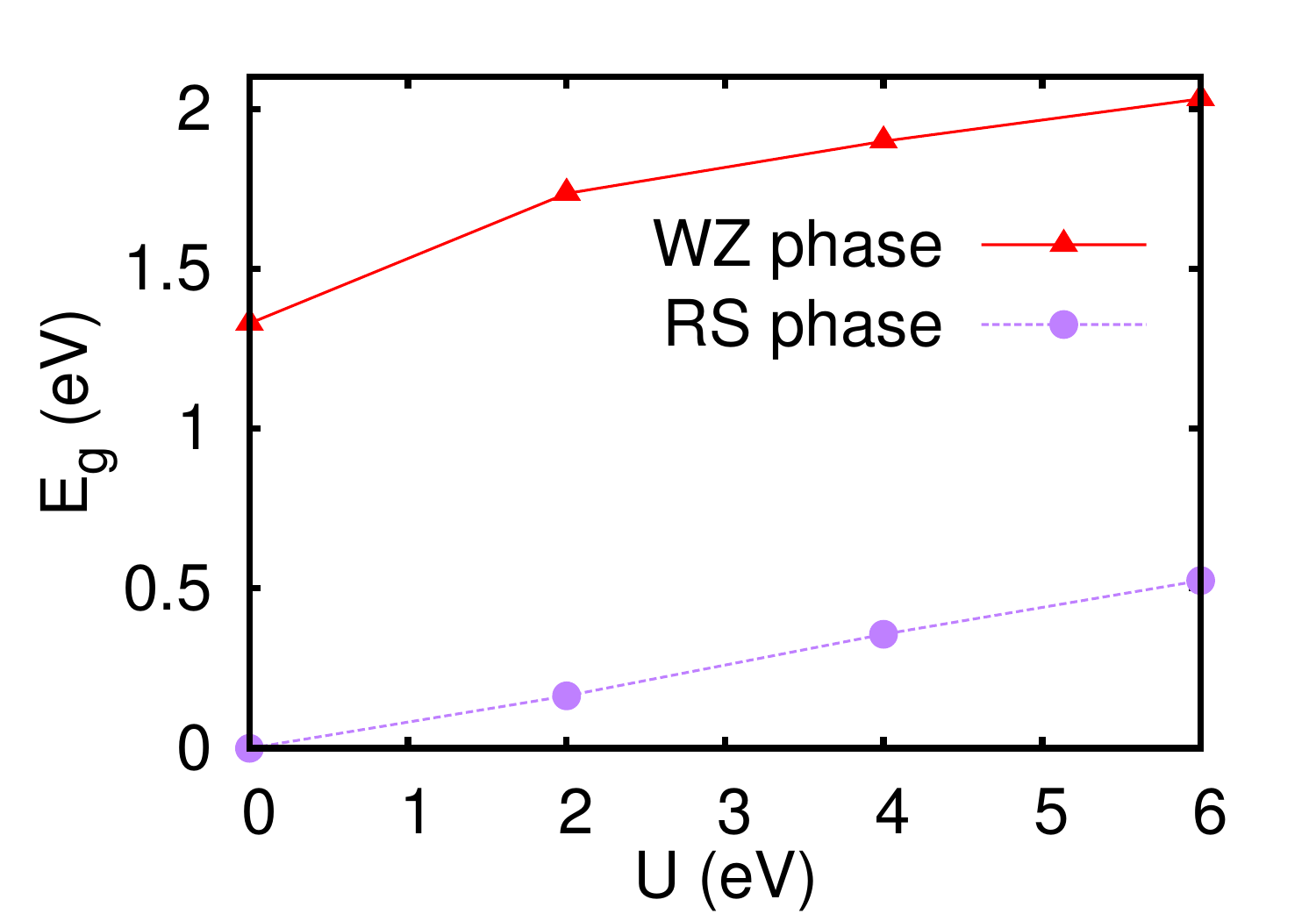}
\caption{Band gap of the rock salt phase (purple line) and wurtzite phase (red line) as a function of the Coulomb repulsion U.}
\label{FigureTheory2}
\end{figure}

Electronic structure calculations were performed within the framework of the first-principles density functional theory based on plane wave basis set and projector augmented wave method using VASP \cite{VASP} package. A plane-wave energy cut-off of 400~eV and the generalised gradient approximation of Perdrew, Burke, and Ernzerhof has been used \cite{perdew1996generalized}. We have performed the calculations using 10$\times$10$\times$6 k-points centered in $\Gamma$. The Hubbard U effects for the 3d-orbitals of Mn$^{2+}$ have been adopted \cite{Liechtenstein95density}. We have scanned the value of U from 0 to 6 eV \cite{Ivanov2016,PhysRevResearch.4.023256} keeping J$_H$ = 0.15U.
Using first-principle calculations, we compare the rock-salt crystal structure (space group no. 225) with the wurtzite (space group no. 186). The band gap changes from indirect in the rock-salt phase\cite{Youn2004} to direct in the wurtzite phase as we can see in Fig. \ref{FigureTheory1}(a,b). 
In the wurtzite phase, the valence band is composed of p-Se from -4 eV up to the Fermi level.  The majority of d-bands are quite flat and present at between -5 and -6 eV as happens in (Cd,Mn)Te \cite{Autieri21}.
At the $\Gamma$ point, the bottom of the conduction is composed of s-bands of Te and Mn.
We investigate the band gap E$_g$ as a function of the Coulomb repulsion U in Fig. \ref{FigureTheory2}. As usual, the band gap increases monotonically as a function of U for both cases and we find that the band gap of the wurtzite is always 1.5 eV larger than the rock-salt phase. We recall that {\it ab initio} calculations tend to underestimate the gap, this underestimation is particularly severe in the WZ phase as we can see from ZnO calculations \cite{HARUN2020102829}. However, we expect that the prediction of the difference between the gaps would be much more reliable, therefore, we expect that the wurtzite band gap is around 1.5 eV larger than the rock-salt phase band gap.

\begin{figure}[th]
\centering
\includegraphics[angle=270,scale=0.34]{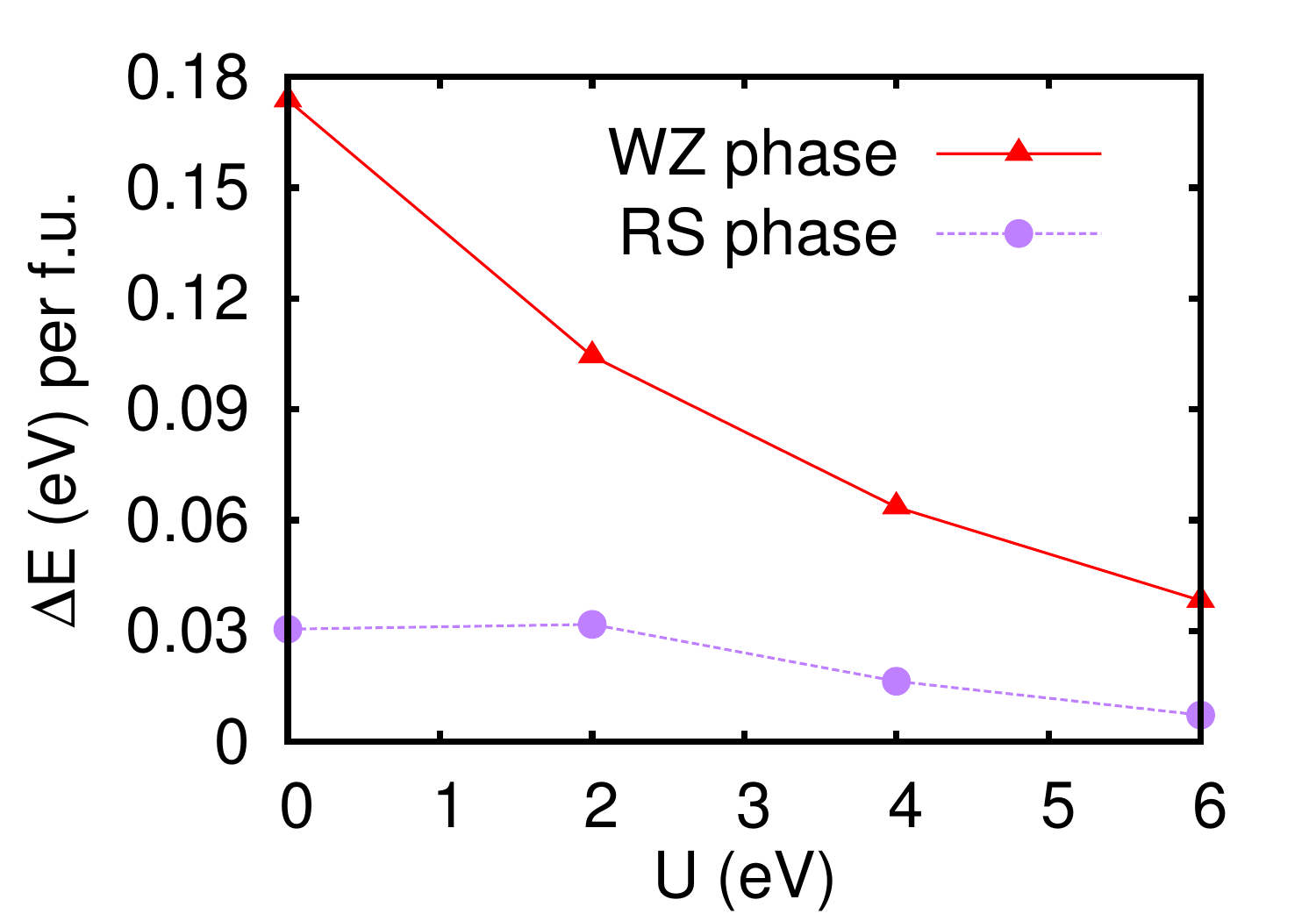}
\caption{Energy difference per formula unit for the rock-salt phase (purple line) and wurtzite phase (red line) as a function of the Coulomb repulsion. The deviation from the trend for the rock-salt phase at low U is due to the insulator-to-metal transition of the FM phase at U=0.}
\label{FigureTheory3}
\end{figure}

To investigate the magnetic properties we analyze the quantity ${\Delta}$E=E$_{FM}$-E$_{AFM}$ per formula unit as a function of U for both rock-salt and wurtzite phases. The quantity ${\Delta}$E is proportional to the magnetic exchange and therefore to T$_N$.
As we can see from Fig. \ref{FigureTheory3}, in both cases, the AFM phase is the ground state. However, in the rock-salt phase, the AFM is magnetically frustrated producing a relatively low value of the ${\Delta}$E while the energy difference is much larger for the wurtzite. Considering that the rock-salt phase has T$_N$=120 K detected experimentally, the wurtzite MnSe has T$_N$ well above room temperature. From {\it ab initio} calculations, the N\'eel vector is in the ab plane with a small energy difference between the magnetic configuration with spins in the plane and out of the plane. The magnetocrystalline anisotropy is of the order of 0.02-0.06 eV per formula unit for values of the Coulomb repulsion U in the physical range of U between 2 and 6 eV.\\

\begin{figure}[th]
\centering
\includegraphics[angle=270,scale=0.34]{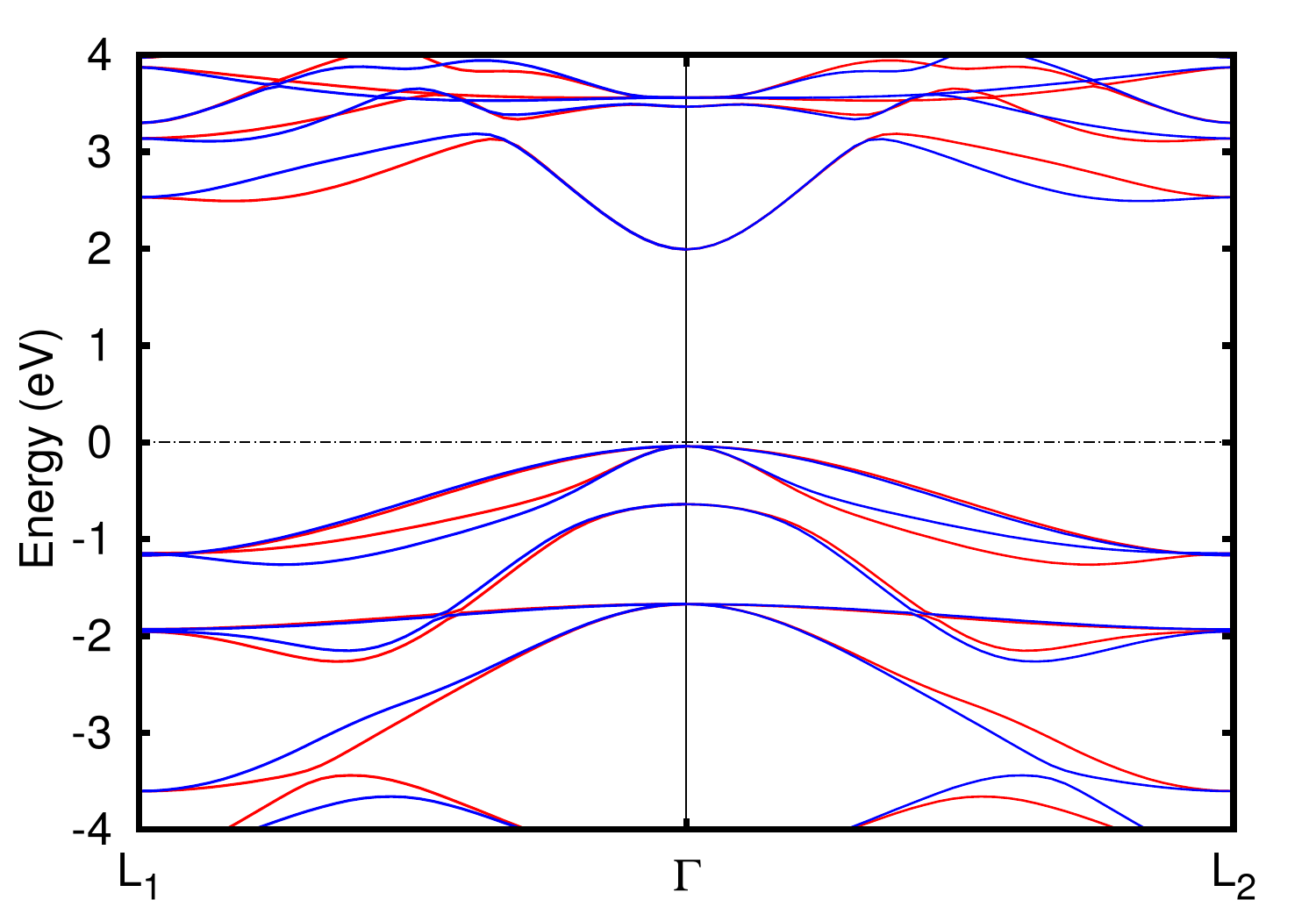}
\caption{Magnetic band structure of wurtzite MnSe along the high-symmetry positions L$_1$-$\Gamma$-L$_2$ in the absence of SOC. Blue and red lines represent the spin-up and spin-down channels, respectively. The Fermi level is set to zero. Altermagnetism is revealed by the lifted degeneracy of the spin-up and spin-down channels.}
\label{FigureTheory4}
\end{figure}

Regarding the altermagnetism, the rock-salt phase is not altermagnetic since both magnetic and non-magnetic atoms are in high-symmetry positions. The altermagnetic properties are present in wurtzite MnSe and similar to the NiAs phase except that the non-relativistic spin-splitting is much smaller in the wurtzite MnSe. The band structure along the k-path that shows altermagnetism is presented in Fig.~\ref{FigureTheory4}. With subscripts 1 and 2 in Fig.~\ref{FigureTheory4}, we indicate the two points in the k-space that have opposite non-relativistic spin-splitting. The points L$_1$ and L$_2$ have the same position of the Brillouin zone as it is in the case of MnTe \cite{Sattigeri23altermagnetic}. 

\section{Conclusions}
In summary, we report on the MBE growth and structural studies of thin layers of 
 semiconductor MnSe in (111) orientation for the first time. We reveal the possibility of controlling the crystal structure of the resulting MnSe. The growth directly on GaAs substrates leads to the typical rock-salt structure of MnSe, whereas the presence of underneath wurtzite CdSe buffer layer makes MnSe adopt the wurtzite structure, which is a rare phase for this compound. Within {\it ab initio} calculations, we have demonstrated that the wurtzite MnSe is an altermagnetic compound with lower non-relativistic spin-splitting with respect to the NiAs-phase of MnTe. Moreover, the calculations show that wurtzite MnSe has a direct band gap of 1.5 eV larger than the rock-salt phase of MnSe and a N\'eel temperature above room temperature. Such interesting observations extend the array of materials available for the physics of altermagnetism and highlight the role of the proper choice of the buffer layers in epitaxial growth of manganese-based chalcogenides. 

\section*{Conflicts of interest}
There are no conflicts to declare.

\section*{Acknowledgements}
This research was funded by the National Science Centre, Poland 2021/40/C/ST3/00168. C. A. is supported by the Foundation for Polish Science through the International Research Agendas program co-financed by the European Union within the Smart Growth Operational Programme (Grant No. MAB/2017/1).
C. A. acknowledges the access to the computing facilities of the Interdisciplinary Center of Modeling at the University of Warsaw, Grant g91-1418, g91-1419 and g91-1426 for the availability of high-performance computing resources and support. C. A. acknowledges the CINECA award under the ISCRA initiative  IsC99 "SILENTS”, IsC105 "SILENTSG" and IsB26 "SHINY" grants for the availability of high-performance computing resources and support. C. A. acknowledges the access to the computing facilities of the Poznan Supercomputing and Networking Center Grant No. 609. J. S. acknowledges a support within "New Ideas 2B in POB II” IDUB project financed by the University of Warsaw.\\

For the purpose of Open Access, the authors have applied a CC-BY public copyright licence to any Author Accepted Manuscript (AAM) version arising from this submission.

%\nocite{*}
\bibliography{wurtzite_vs_rocksalt_MnSe_arxiv}% Produces the bibliography via BibTeX.

\end{document}